\documentclass[smallextended]{svjour3}       
\smartqed  

\usepackage{amsmath,amssymb}
\usepackage{graphicx}
\usepackage{gensymb}
\usepackage{xcolor}
\usepackage{soul}

\journalname{Eur. Phys. J. E}

\newcommand{\mean}[1]{\langle #1\rangle}
\newcommand{\pd}[2]{\frac{\partial #1}{\partial #2}}
\newcommand{\is}{\text{Ising}}

\begin{document}

\title{Critical behaviour in active lattice models of motility-induced phase separation}

\author{Florian Dittrich            \and
        Thomas Speck                \and
        Peter Virnau
}

\institute{Institute of Physics, Johannes Gutenberg-Universität, Mainz, Germany \\
              \email{virnau@uni-mainz.de}
}

\date{Received: date / Accepted: date}

\maketitle

\begin{abstract}
Lattice models allow for a computationally efficient investigation of motility-induced phase separation (MIPS) compared to off-lattice systems. Simulations are less demanding and thus bigger systems can be accessed with higher accuracy and better statistics. In equilibrium, lattice and off-lattice models with comparable interactions belong to the same universality class. Whether concepts of universality also hold for active particles is still a controversial and open question. Here, we examine two recently proposed active lattice systems that undergo MIPS and investigate numerically their critical behaviour. In particular, we examine the claim that these systems and MIPS in general belong to the Ising universality class. We also take a more detailed look on the influence and role of rotational diffusion and active velocity in these systems.
\end{abstract}


\section{Introduction}
\label{intro}

Non-equilibrium active systems composed of self-propelled particles offer a wide range of interesting behaviour and applications~\cite{Vicsek:2012,Bechinger:2016,roadmap}. A fundamental phenomenon is the so called motility-induced phase separation (MIPS)~\cite{Cates:2015}: At large propulsion speeds and low rotational diffusion, self-propelled particles block each other due to excluded volume and form initial clusters. If the time-scale for the rotational diffusion of the particle orientations at the border of such a cluster is larger than the time it takes to enrich the cluster with additional particles, a dynamical instability leading to non-equilibrium phase separation is induced. Although the phase-separated state resembles passive liquid-gas separation with dense domains surrounded by an active gas, no explicit attractive interactions are required. Still, the phase behavior is very similar, with the binodal curve of coexisting densities terminated by a critical point below which the system remains homogeneous for all densities. The question whether or not the behaviour close to such a non-equilibrium critical point is universal, and if it can be attributed to one of the standard universality classes, is not only of fundamental interest but has also stirred up an ongoing controversy~\cite{Siebert:2018,Caballero:2018,Partridge:2019,Maggi:2020}.

Numerical studies of active Brownian particles (ABPs)~\cite{Fily:2012,Redner:2013,Stenhammar:2013,Stenhammar:2014,Wysocki:2014,Bialke:2015,Siebert:2017,Digregorio:2018} are a common approach to investigate MIPS in a simple continuous system. These particles are modelled as disks interacting with each other via a purely repulsive Weeks-Chandler-Anderson potential in the framework of an overdamped Langevin equation. In addition, they are propelled with constant speed along their orientation, which is subject to rotational diffusion. For this system, we have determined the location of the critical point in two dimensions and reported critical exponents, which are incompatible with any of the known universality classes~\cite{Siebert:2018}. To gain access to the critical point and the critical exponents, we have proposed a novel method to sample subboxes that minimizes the influence of interfaces on density fluctuations.

In contrast, subsequent numerical investigations of related but different models have come to a different conclusion, supporting Ising universality in two dimensions for off-lattice Active Ornstein-Uhlenbeck particles~\cite{Maggi:2020} and a lattice variant of ABPs~\cite{Partridge:2019} (and for ABPs in three dimensions~\cite{Turci:2020,Omar:2020}). Following generic arguments of renormalization~\cite{Hohenberg:1977}, however, all these models (in two dimensions) should fall into the same universality class and thus exhibit the same critical exponents. Indeed, in a first renormalization study of an active field theory (``active model B+''~\cite{Nardini:2017,Tjhung:2018}) it was found that the critical behavior is controlled by the Wilson-Fisher fixed point~\cite{Caballero:2018}. What, then, is the reason for the reported differences? Regarding geometry and subboxes, all three numerical works have employed the same method (for details, see Sec.~\ref{sec:method}). One reason could be insufficient statistics, or insufficient range of system sizes leading to a biased estimate of critical exponents. Or, more intriguingly, are there additional features that characterize universality classes in active matter? We stress that MIPS of repulsive particles is a genuine non-equilibrium phenomenon. The effect of self-propulsion on the critical behavior in models that exhibit phase separation already under equilibrium conditions has been studied for Lennard-Jones interactions~\cite{Prymidis:2016} and a three-dimensional Asakura-Oosawa model~\cite{Zausch:2009} driven by a Vicsek-type force~\cite{Vicsek:1995,Das:2014,Trefz:2016} and found to be compatible with the 3d-Ising universality class~\cite{Trefz:2017}.

In this manuscript, we take another step towards a comprehensive understanding of critical behaviour in active matter. To this end, we numerically investigate different variants of two-dimensional active lattice gases with excluded volume and dynamics that break detailed balance. While a range of active lattice gas models has been investigated~\cite{Thompson:2011,Soto:2014,Pilkiewicz:2014,Solon:2015a,Manacorda:2017,Houssene:2018}, we focus on two variants that mimic active Brownian particles. In particular, we study two lattice geometries (the square and hexagonal lattice) and two implementations of the dynamics, either treating rotation and translation serial~\cite{Partridge:2019} or concurrently~\cite{Whitelam:2018}. Our numerical results indicate that details of the dynamics have an influence on the critical behavior and question the proposition that MIPS falls into Ising universality.


\section{Methods}

\subsection{Critical behavior}

Before embarking on the computational study, let us recall some of the properties close to a critical point. We consider systems that undergo phase separation with two coexisting phases having different densities $\rho$. The two phases are identified with gas ($\rho_\text{gas}$) and liquid ($\rho_\text{liq}$). The (average) order parameter is the difference, $m=\rho_\text{liq}-\rho_\text{gas}$. As we approach the critical point, the gap in $m$ closes and follows a path through the critical point. Hence, we observe
\begin{equation}
    m \sim \tau^\beta,
\end{equation}
whereby $\tau$ measures the distance to the critical point (typically the reduced temperature) and $\beta$ is the corresponding critical exponent. The transition is continuous, with $m>0$ for $\tau>0$ and $m=0$ in the homogeneous phase for $\tau<0$. In addition, both the susceptibility $\chi$ and the correlation length $\xi$ diverge at the critical point,
\begin{equation}
    \chi \sim \tau^{-\gamma}, \qquad \xi \sim \tau^{-\nu},
\end{equation}
defining two more exponents. Of particular importance is Ising universality in equilibrium systems with short-range interactions and scalar order parameter, for which in two dimensions the exponents can be obtained analytically~\cite{Onsager:1944,Yang:1952,Wu:1976}
\begin{equation}
    \beta = \tfrac{1}{8}, \qquad \gamma = \tfrac{7}{4}, \qquad \nu = 1.
    \label{eq:ising}
\end{equation}
Note that these three critical exponents are not independent but obey the \emph{hyperscaling relation}
\begin{equation}
    \gamma + 2\beta = 2\nu.
    \label{eq:hyper}
\end{equation}
Arguments why this relation might still be valid for driven active systems are sketched in appendix~\ref{sec:hyper}.

The diverging correlation length $\xi$ implies that the critical behavior is modified in finite systems, where the correlation length is bounded by the system size $l$. One of the most remarkable successes of computational statistical physics is that the critical behavior can still be extracted from simulations of finite systems~\cite{Binder:1981,Binder:1987}. To locate the critical point, we turn to Binder's cumulant ratio
\begin{equation}
    Q_l(\tau) = \frac{\langle m_l^2\rangle^2}{\langle m_l^4\rangle},
    \label{eq:binder}
\end{equation}
which becomes independent of $l$ exactly at the critical point. Note that $m_l=(N_l-\mean{N_l})/l^2$ in the lattice gas formulation with $N_l$ being the number of particles. Plotting the ratio $Q_l$ as a function of some parameter for different $l$ thus allows--notwithstanding systematic effects as discussed below--to locate the critical point from the intersection of curves. Moreover, the derivative $dQ_l/d\tau|_{\tau=0}=1/\nu$ yields the inverse of the critical exponent $\nu$. Once we have located the critical point, we can extract $\beta$ from plotting $\langle m_l\rangle$ as a function of $\tau$. Finally, we exploit the scaling form $\chi_l=l^{\gamma/\nu}\tilde\chi(l/\xi)$ for the susceptibility with scaling function $\tilde\chi$ that depends on system size through the ratio $l/\xi$. Plotting the susceptibility (obtained from the fluctuations of the order parameter) as a function of $l$ allows to extract the ratio $\gamma/\nu$. We thus have access to the three critical exponents $\nu$, $\gamma$, and $\beta$.

\subsection{Simulations}
\label{sec:method}

While the ensemble of choice for simulations of critical behavior in equilibrium is the grand-canonical ensemble, for driven active systems breaking detailed balance this route is not available due to the absence of a comprehensive framework in which a chemical potential is defined (although attempts have been made~\cite{Paliwal:2018,Solon:2018}).

\begin{figure}[b!]
\begin{centering}
\includegraphics[width=0.6\textwidth]{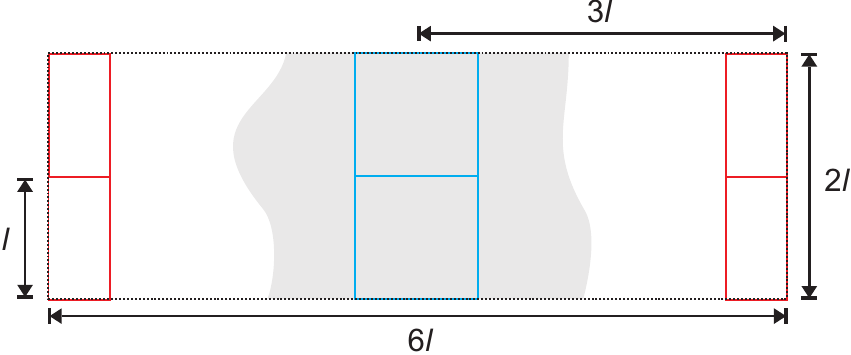}
\caption{Sketch to illustrate the simulation box setup. The evaluated subboxes of size $l\times l$ are placed in the dense (cyan) and dilute (red) phase. The overall simulation box is set to size $2l\times 6l$. }
\label{fig:meth}
\end{centering}
\end{figure}

Therefore, we closely follow the method and analysis proposed in Ref.~\cite{Siebert:2018}. All simulations were performed in a periodic box with 1:3 geometry. In such elongated boxes, the dense phase nucleates to a slab-like structure, cf. Fig.~\ref{fig:meth}. The slab arises along the short side of the box and connects to itself via periodic boundary conditions. Its position inside the simulation box can be easily determined as the center of mass of all particles.

In order to measure dense and dilute phase, as well as to avoid interface regions between the two phases, we place two quadratic subboxes of size $l$ above each other right in the center of mass to sample the dense phase (see Fig. \ref{fig:meth}). Another set of two subboxes is placed shifted away by one half of the simulation box's width from the center of mass to sample the dilute phase. In total, we sample the number of particles $N_l$ within the four subboxes of size $l$ in a simulation box of total size $2l\times 6l$. Note that $N_l$ is counted for each of the four subboxes separately, resulting in four measurements for each snapshot. By adjusting the size of the simulation box through the subbox size, we couple the maximum correlation length to $l$ and achieve a clear crossing point of $Q_{l}$ for different $l$. The susceptibility is evaluated as $\chi_{l}=\mean{(N_l-\mean{N_l})^2}/\mean{N_l}$. Coexistence densities of dense and dilute phase ($\rho_{\text{liq}}$ and $\rho_{\text{gas}}$) are obtained as plateau values of density profiles generated from a simulation box of size $252\times 84$ (corresponding to $l = 42$) at activities slightly above the tentative critical point.

We use the same system sizes of $l=12,18,24,30,36,42$ for all simulations, resulting in simulation boxes of size $24\times 72$, $36\times 108$, $48\times 144$, $60\times 180$, $72\times 216$, $84\times 252$. The number density is always 0.5, which gives the corresponding particle numbers 864, 1944, 3456, 5400, 7776, 10584. Furthermore, we chose all activities to be at comparable relative distance to the critical point.

\begin{figure}[t]
  \includegraphics[width=1.0\textwidth]{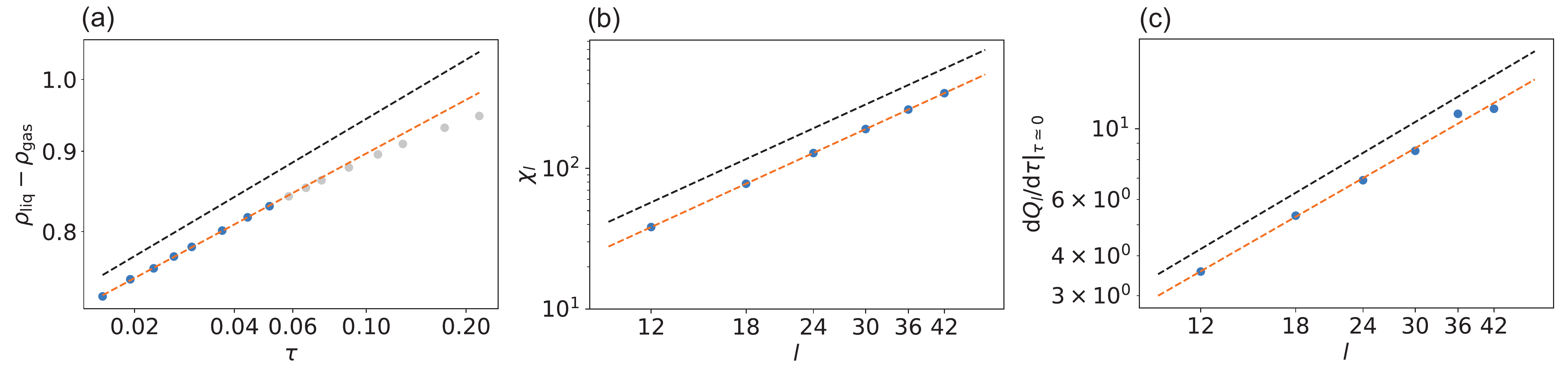}
  \caption{Estimation of the critical exponents for the 2d Ising model. The black dashed lines show the slopes with the analytical values [Eq.~\eqref{eq:ising}]. Grey dots in (a) were excluded from the analysis since they are too far from the critical point and have moved away from the power-law scaling.}
  \label{fig:ising_crex}
\end{figure}

As reference, in Fig.~\ref{fig:ising_crex} we show our analysis applied to the 2d Ising model. Since the 2d Ising model is subject to critical slowing down, we get somewhat poorer statistics (see Appendix \ref{sec:slowing}). Nevertheless, the analysis yields the following critical exponents
\begin{equation}
  \beta \simeq 0.113(1), \qquad \gamma/\nu \simeq 1.751(2), \qquad 1/\nu \simeq 0.97(3)
\end{equation}
in good agreement with the analytical values. Errors in this and later sections refer to statistical errors obtained by splitting respective data sets into three parts and calculating the standard error of the mean. While $\gamma/\nu$ agrees exactly, there is a slight underestimation for $1/\nu$ and a noticeable underestimation for $\beta$. While the error of $1/\nu$ might be within statistical uncertainties, the deviation observed for $\beta$ is more systematic: The power law scaling is clearly only valid very close to the critical point. However, measuring points closer to the critical point than shown in Fig.~\ref{fig:ising_crex} is challenging. Below we show that the active lattice models follow a power-law behavior over a wider range.


\section{Model I: Serial rotation/translation}

\subsection{Model description}

\begin{figure}[t]
  \includegraphics[width=1.0\textwidth]{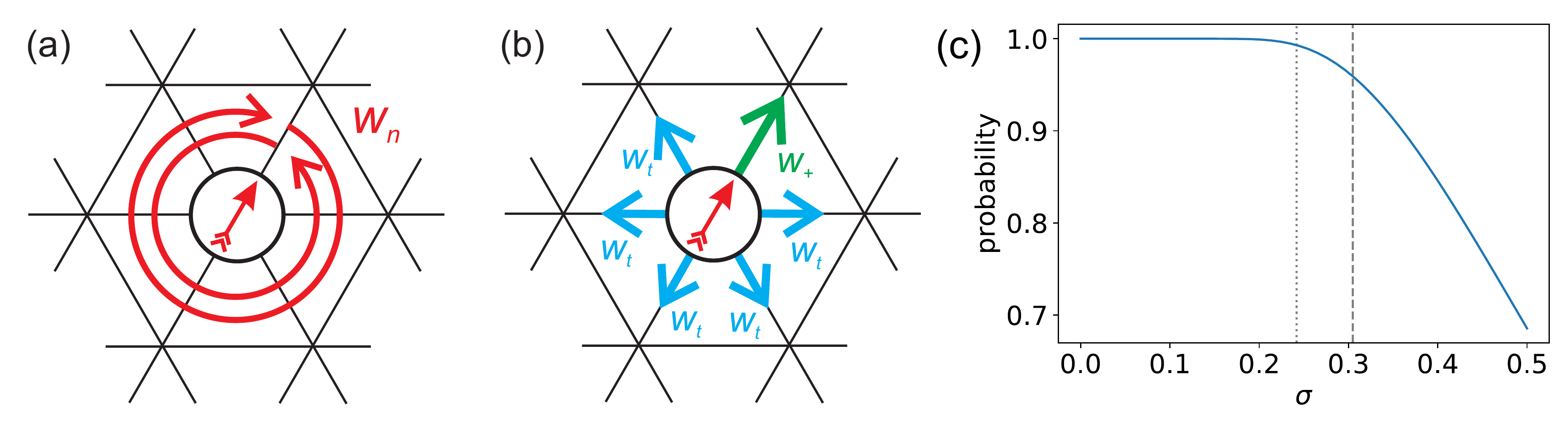}
  \caption{Sketch to illustrate serial model I on a hexagonal lattice. (a)~First, the orientation (arrow) of the particle is updated drawing a random number from a Gaussian distribution. (b)~Then, a move along the particle's orientation is attempted with rate $w_+$, diffusion along another direction is attempted with rate $w_t$ each. (c)~Probability to keep the current orientation as a function of $\sigma$ (as given by the integral over the Gaussian distribution from $-0.5$ to $0.5$). Dashed and dotted lines correspond to critical values for model I on the hexagonal ($\sigma_\text{c}\simeq0.3048$) and square lattice ($\sigma_\text{c}\simeq0.2415$).}
  \label{fig:hlm}
\end{figure}

We first turn to the model studied in Ref.~\cite{Partridge:2019} employing a hexagonal lattice, which we will refer to as model I. On a hexagonal lattice each particle has six neighbouring sites and six discrete directions it can be orientated towards (Fig.~\ref{fig:hlm}). Specifically, each Monte Carlo (MC) step works as follows:
\renewcommand{\labelenumi}{\arabic{enumi}.}
\begin{enumerate}
  \item A particle is picked at random.
  \item A Gaussian distributed random number (with standard deviation $\sigma$ and zero mean) is drawn and rounded to the nearest integer $n$. The current orientation of the particle is adjusted by that integer ($n=1$ means one step clockwise, $n=-1$ means one step counterclockwise, $n=2$ means two steps clockwise, and so on), cf. Fig.~\ref{fig:hlm}(a).
  \item A movement along the new orientation of the particle is chosen with probability $w_+=25/30$, other directions are chosen with probability $w_t=1/30$ each mimicking translational diffusion, cf. Fig.~\ref{fig:hlm}(b).
  \item If the target lattice site is empty the move is accepted, otherwise the move is rejected. This ensures that each lattice site is either unoccupied or occupied by exactly one particle.
\end{enumerate}

Note that the adjustment of orientation in step 2 is always accepted and a translation does not change the orientation of the particle. Since $w_+$ and $w_t$ are fixed, the ``activity'' of the system is solely adjusted via the rotational diffusion, which is defined by the width $\sigma$ of the Gaussian distribution. A low value for $\sigma$ corresponds to low rotational diffusion and therefore highly persistent motion [see Fig.~\ref{fig:hlm}(c)]. Note that the probability to keep the current orientation is not linear in $\sigma$, especially not around the estimated values for the critical points. It is also important to note that in contrast to Model II discussed below, rotation (step 2) and translation (step 4) are always performed in series.

\subsection{Analysis and results}

\begin{figure}[t]
  \centering
  \includegraphics[width=0.9\textwidth]{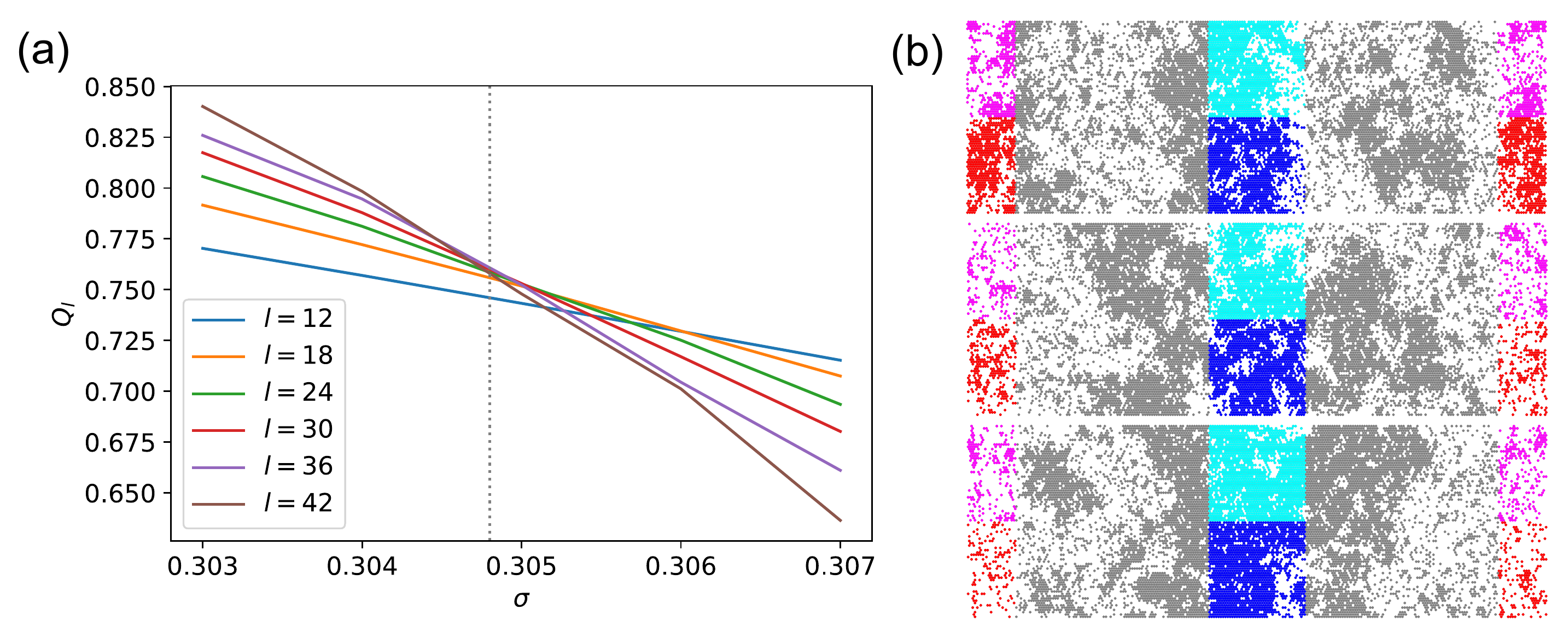}
  \caption{(a) Cumulant ratios $Q_l$ for model I on the hexagonal lattice. The dotted line indicates the estimated critical value $\sigma_c\simeq0.3048$ as the mean crossing point for $l=24,30,36,42$ excluding the two smallest system sizes. Note that each tick on the $x$-axis corresponds to one simulation point. (b) Snapshots for the largest system of $l=42$ below the critical point at $\sigma=0.307$ (top), at the critical point at $\sigma=0.3048$ (middle) and above the critical point at $\sigma=0.303$ (bottom). Particles are colored according to the subbox they are in.}
  \label{fig:hlm_cc}
\end{figure}

By closely following the analysis described in Sec.~\ref{sec:method}, we determine the critical point $\sigma_{cr,I}\simeq0.3048$ as the average of the cumulant ratio crossings (Fig.~\ref{fig:hlm_cc}) for the four largest system sizes under consideration ($l=24,30,36,42$). This value is in agreement with the results published in Ref.~\cite{Partridge:2019}, which has analyzed systems of comparable system sizes. Excluding the two smallest boxes ($l=12,18$)\footnote{The fact that only intermediate box sizes cross is common for block-density-distribution methods also in passive systems~\cite{Watanabe:2012}.}, the cumulant ratios for the bigger boxes cross within a small interval as expected for critical scaling.

\begin{figure}[t]
  \includegraphics[width=1.0\textwidth]{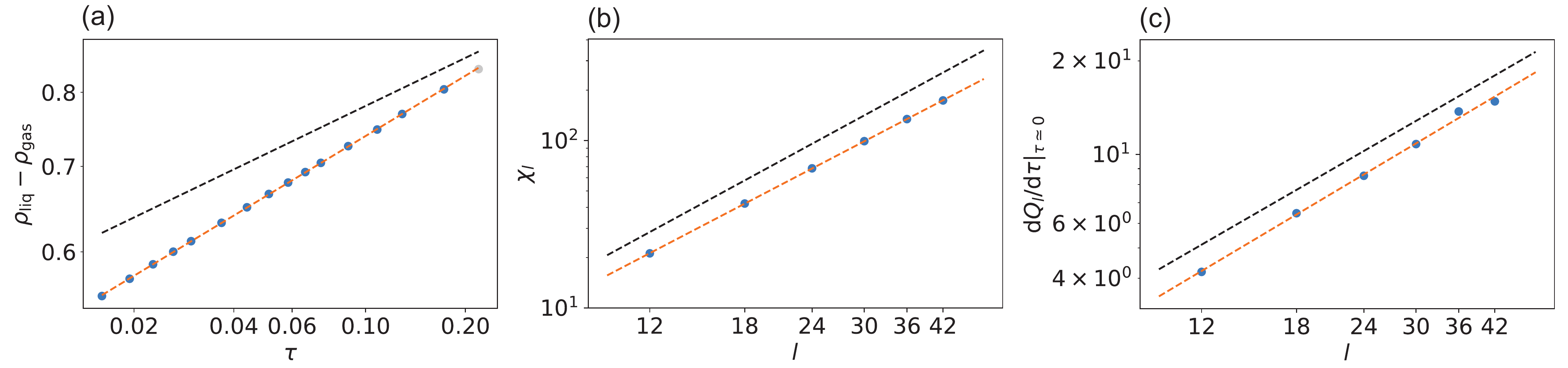}
  \includegraphics[width=1.0\textwidth]{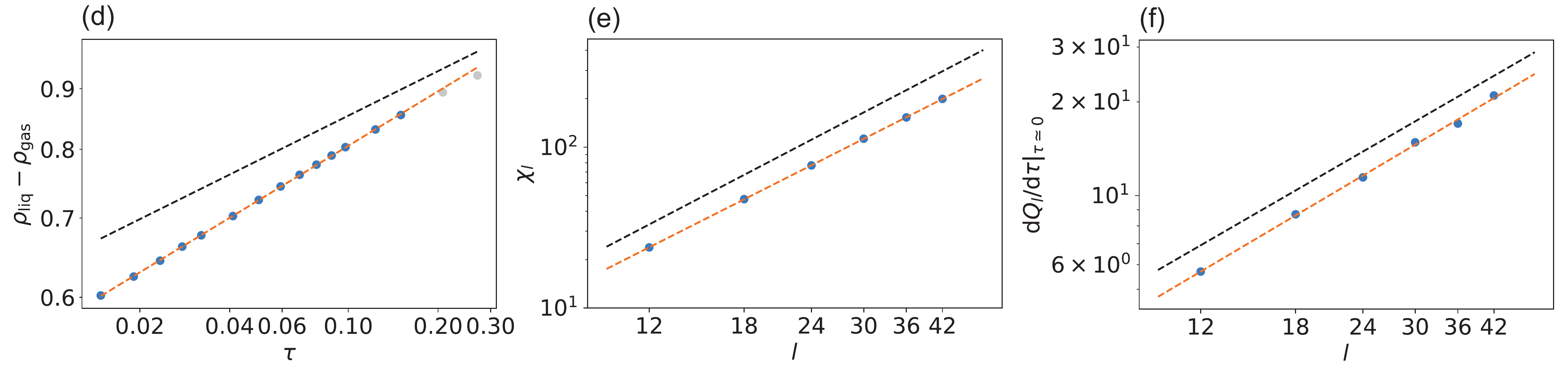}
  \caption{(a-c) Estimating the critical exponents for model I. Plotted are (a)~the order parameter $\rho_\text{liq}-\rho_\text{gas}$, (b)~the susceptibility $\chi_l$, and (c)~the slope of the cumulant ratio at the critical point. (d-f) Corresponding determination for model I but on a square lattice with $\sigma_c\simeq0.2415$. Grey dots in (a) and (d) were excluded from the analysis. The black dashed lines show the slopes with the critical exponents for the 2d Ising system.}
  \label{fig:hlm_crex}
\end{figure}

Fig.~\ref{fig:hlm_crex}(a-c) shows results for the order parameter, the susceptibility, and the derivative of the cumulant ratio. Fitting power laws yields the following exponents
\begin{equation}
  \beta \simeq 0.1567(3), \qquad \gamma/\nu \simeq 1.678(2), \qquad 1/\nu \simeq 1.03(2)
\end{equation}
and thus $\nu\simeq0.97$ and $\gamma\simeq1.63$. While the agreement with the corresponding 2d Ising values is reasonable for $\nu$ ($\nu_\is=1.0$) and $\gamma$ ($\gamma_\is=1.75$), the exponent $\beta$ differs by more than $25\%$ from $\beta_\is=0.125$. This disagreement is also clearly visible in Fig.~\ref{fig:hlm_crex}(a).

To test the influence of the underlying lattice geometry, we have also performed an analogous investigation of the model on a square lattice with $w_+=17/20$ for movements along the particles current orientation and $w_t=1/20$ for the three remaining directions. The results are shown in Fig.~\ref{fig:hlm_crex}(d-f) with exponents
\begin{equation}
    \beta \simeq 0.1528(1), \qquad \gamma/\nu \simeq 1.695(3), \qquad 1/\nu \simeq 1.023(8)
\end{equation}
and thus $\nu\simeq0.98$ and $\gamma\simeq1.66$. These critical exponents are very similar to the hexagonal case and within numerical uncertainties, indicating that the influence of the underlying lattice is negligible as one would expect.

\section{Model II: Concurrent rotation/translation}

\subsection{Model description}

\begin{figure}[b!]
  \centering
  \includegraphics[width=0.57\textwidth]{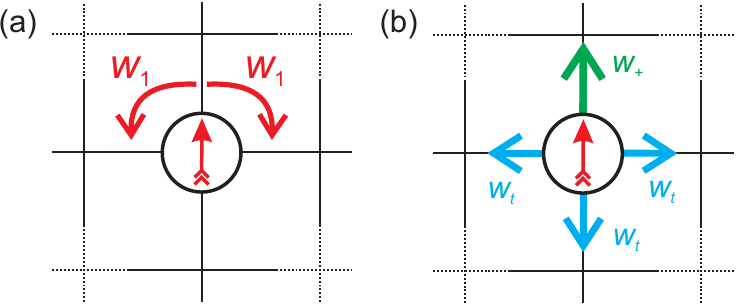}
  \caption{Sketch to illustrate model II on a square lattice. (a) The particle orientation is turned clockwise or counterclockwise by $90\degree$ with rate $w_1$. (b) A move along the orientation is attempted with rate $w_+$, diffusion into any other direction with rate $w_t$ each. Note that only one of these moves is attempted in each time step. The probability for each move is given by the respective rate divided by the sum of all rates.}
  \label{fig:sqlm}
\end{figure}

The second model is based on a square lattice and has been proposed in Ref.~\cite{Whitelam:2018}. As illustrated in Fig.~\ref{fig:sqlm}, there are now six possible moves: either rotation of the particle orientation clockwise or counterclockwise by $90\degree$ with weight $w_1$ [Fig.~\ref{fig:sqlm}(a)], or translation along the orientation with weight $w_+$ or any of the three other directions with weight $w_t$ [Fig.~\ref{fig:sqlm}(b)]. In contrast to Model I, the weight $w_1=0.1$ for rotation is now kept constant and we vary $w_+$ with $w_t=1$. Moreover, in each MC step one of the moves is selected according to its weight. Hence, the particle can either rotate or move in one time step, which we term concurrent.

\subsection{Analysis and results}

\begin{figure}[t]
  \centering
  \includegraphics[width=0.85\textwidth]{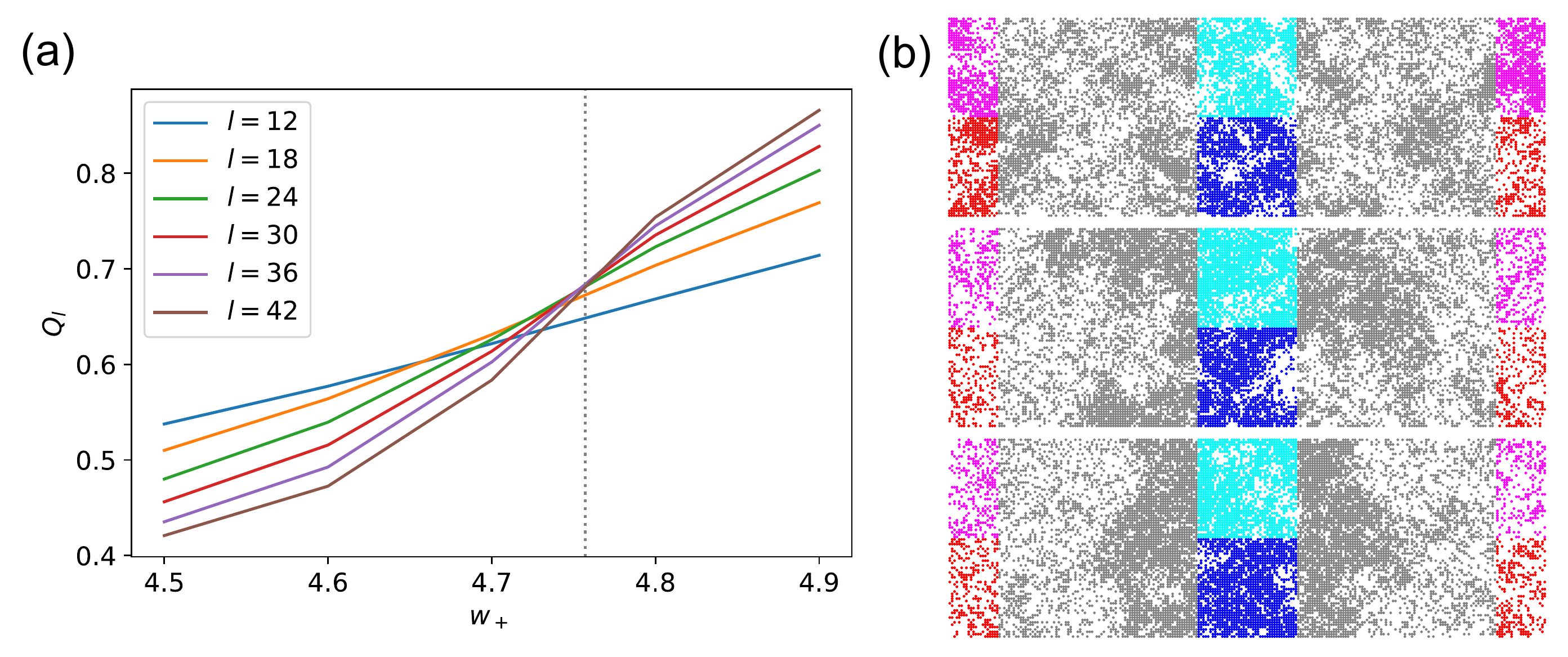}
  \caption{(a) Cumulant crossing for Model II yielding a critical point at $w_{+,\text{cr}}\simeq4.76$ as determined by the mean crossing point for $=24,30,36,42$ (dotted line). Note that each tick on the x-axis represents a $w_{+}$ at which simulations for the various system sizes took place. (b) Snapshots for the largest system of $l=42$ below the critical point at $w_{+}=4.5$  (top), at the critical point at $w_{+}=4.76$ (middle) and above the critical point at $w_{+}=4.9$  (bottom). Particles are colored according to the subbox they are in.}
  \label{fig:sqlm_cc}
\end{figure}

\begin{figure}[t]
  \includegraphics[width=1.0\textwidth]{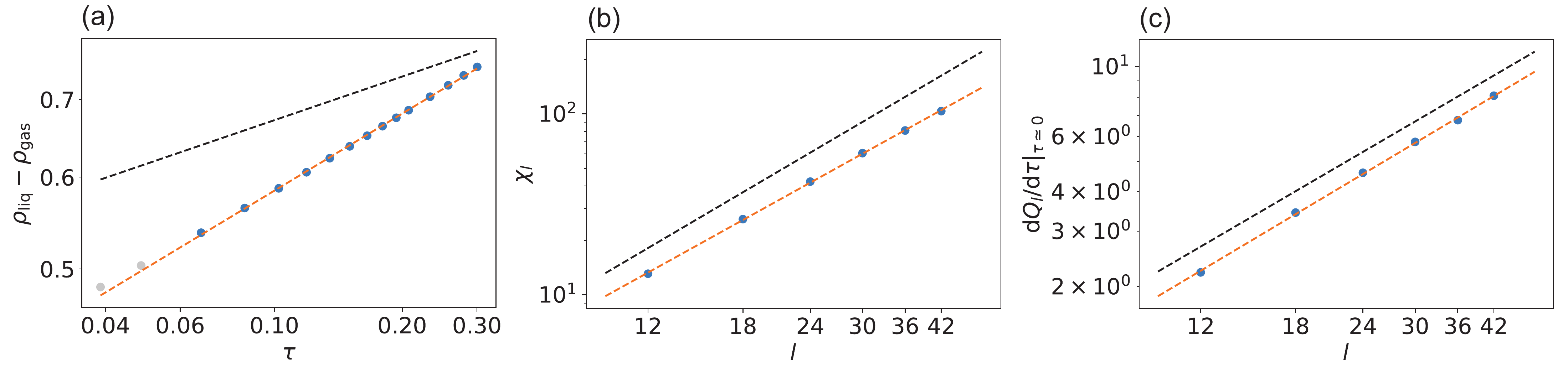}
  \caption{Critical exponents for model II, cf. Fig.~\ref{fig:hlm_crex}. Note that for (a) the grey points were not considered for the fit as slabs started to dissolve, resulting in an ill-defined plateau of the density profiles. In (b) $w_{+}=4.76$ has been used. For comparison, the black dashed lines indicate the 2d Ising critical exponents.}
  \label{fig:sqlm_crex}
\end{figure}

Fig.~\ref{fig:sqlm_cc} shows the crossings of the cumulant ratios $Q_{l}$ for different box lengths $l$. The crossings start to converge for the bigger boxes with $l\geq24$. Hence, we only take these system sizes into account and determine the critical point to be at $w_{+,\text{cr}}\simeq4.76$. Corresponding results for the critical exponents are displayed in Fig.~\ref{fig:sqlm_crex}, for which we find
\begin{equation}
  \beta \simeq 0.2208(1), \qquad \gamma/\nu \simeq 1.649(1), \qquad 1/\nu \simeq 1.021(7).
\end{equation}
While $\gamma\simeq1.68$ and $\nu\simeq0.98$ again exhibit reasonable agreement with 2d Ising values (1.75 and 1, respectively), $\beta=0.221$ exceeds the corresponding value (0.125) by almost a factor of two. Note that $\beta$ needs to be measured further from the critical point than in model I because the density profiles lose their stability faster. This indicates that fluctuations are stronger and the slab in the 1:3 simulation box stays less stable in the vicinity of the critical point for model II.

\section{Alternative determination of $\beta$}
\label{sec:alt_bet}

\begin{figure}[ht]
  \centering
  \includegraphics[width=0.8\textwidth]{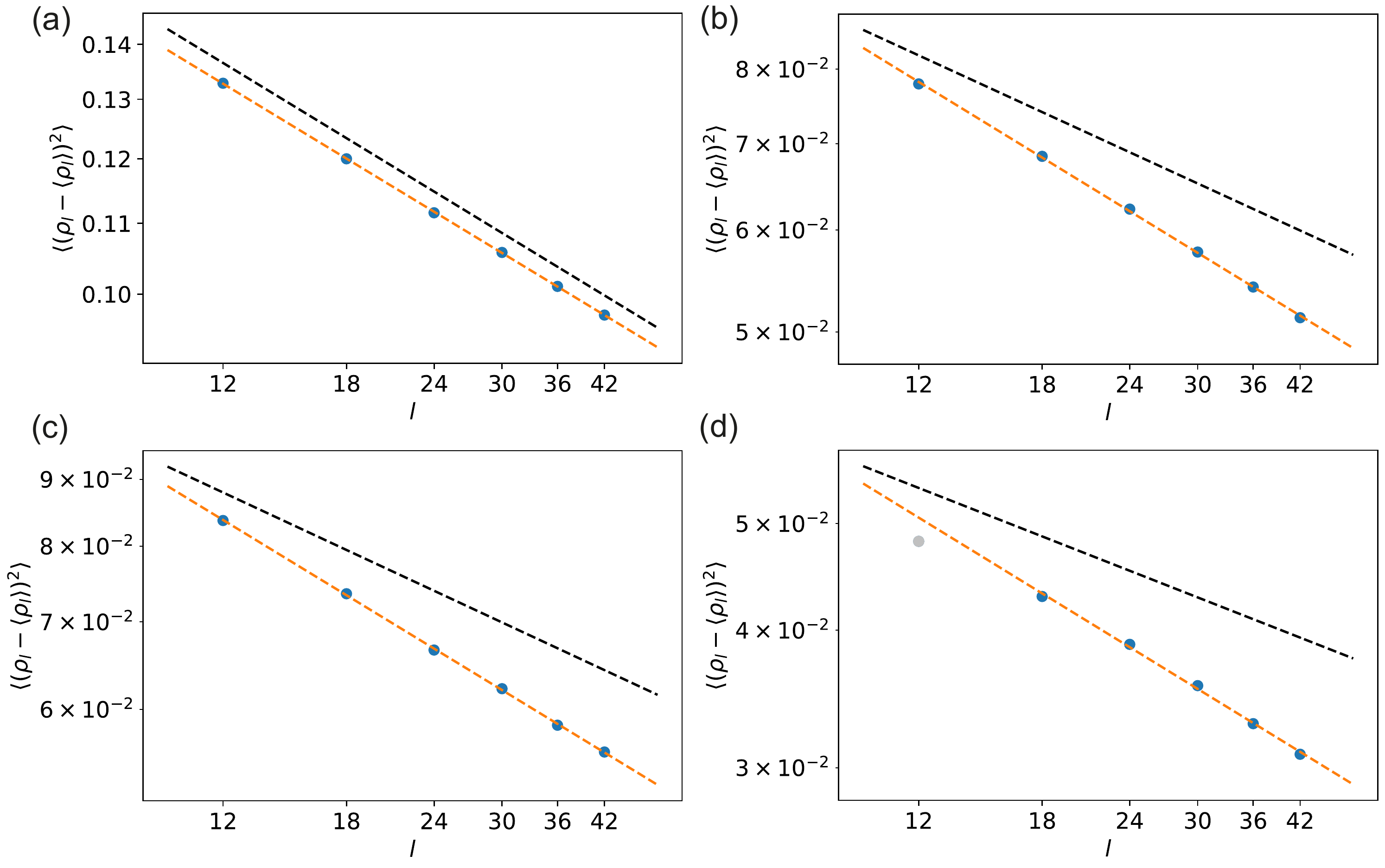}
  \caption{Alternative determination of $\beta$. Plotted is the slope of $\langle(\rho_l-\langle\rho_l\rangle)^2\rangle$ for the different system sizes at the critical point, which yields $-2\beta'/\nu$. For comparison, the black dashed lines indicate the analytical 2d Ising slope of $-1/4$. Fits are shown for: (a) 2d Ising model, (b) model I on hexagonal lattice, (c) model I on square lattice and (d) model II. For model II, there is a noticeable bending and the gray dot in (d) was excluded from the analysis.}
  \label{fig:alt_bet_plot}
\end{figure}

The accuracy of measuring the exponent $\beta$ from the density difference is limited by the fact that it becomes more and more difficult to reliably estimate this difference as we approach the critical point. For the Ising model (cf. Fig.~\ref{fig:ising_crex}), this has led to a noticeable deviation from the known analytical value. There is an alternative method using the density fluctuations $\langle(\rho_l-\langle\rho_l\rangle)^2\rangle \sim l^{-2\beta'/\nu}$ at the critical point which is equivalent to (and therefore not independent from) how $\gamma/\nu$ is determined from $\chi_l$, see also Ref.~\cite{Maggi:2020}. The densities for the subboxes are given by $\rho_l=N_l/l^2$. This analysis is shown in Fig.~\ref{fig:alt_bet_plot}. For the 2d Ising model we now obtain $2\beta'/\nu\simeq0.249(1)$, which is in excellent agreement with the analytical value. The values for the other models are included in Table~\ref{tab:summary}, but it is clear that they substantially deviate from 2d Ising universality.

\section{Discussion and Conclusions}
\label{sec:disc}

Our results for the critical exponents are summarized in Table~\ref{tab:summary}. We have also added the corresponding values for active Brownian particles as determined in Ref.~\cite{Siebert:2018}. For all lattice models studied here, we find values for $\nu$ that are in very good agreement with Ising universality ($<3\%$ smaller) and values for $\gamma/\nu$ that are in good agreement ($<5\%$ smaller). These values are in agreement with plots shown in Ref.~\cite{Partridge:2019}, which concludes that Ising universality holds. This conclusion seems questionable when taking the exponent $\beta$ into account, which deviates substantially. Indeed, the determination of $\beta$ is technically the most challenging. However, note that the hyperscaling relation Eq.~\eqref{eq:hyper} places a strong constraint on the exponents. From $\gamma/\nu\simeq1.68$ and $\nu\simeq0.98$ we can obtain an estimate for $\beta\simeq0.157$ that is in excellent agreement with the numerically estimated values for Model I on both lattice geometries, supporting that reduction of $\gamma$ (and $\nu$) is not a statistical effect but systematic.

The value for $\beta$ estimated for Model II is even larger. However, in this case the hyperscaling relation is only fulfilled approximately, which might indicate that $\beta$ is too large. We have observed that obtaining ``good'' crossings of the cumulant ratio in this model is more challenging, which might be because the speed is changed in contrast to the rotational diffusion in model I. Moreover, the distance to the critical point is larger since the determination of $\rho_\text{liq}-\rho_\text{gas}$ requires stable liquid slabs with a well defined plateau of the density profile. The alternative method of Sec.~\ref{sec:alt_bet} yields a smaller $\beta'\simeq0.19$. We notice that the ratio $\gamma/\nu$ has become even smaller, moving away from the Ising value. While this seems to indicate an influence of the different dynamic rules on the critical behavior, we cannot rule out that the scaling closer to the critical point changes (but note that the smaller value of $\gamma/\nu$ accommodates a larger $\beta$). Even further from Ising universality are off-lattice active Brownian particles, where also the exponent $\nu$ now changes substantially from $\nu=1$ to $\nu\simeq1.5$. Still, the hyperscaling relation is again approximately fulfilled, indicating that the exponents are consistent. We cannot exclude the possibility that corrections to scaling are relevant and modify these exponents in a way that is compatible with scaling relations~\cite{Aharony:1980}.

\begin{table}[b!]
    \centering
    \begin{tabular}{c|c|c|c|c|c}
    \hline\hline
    model & $\beta$ & $\gamma / \nu$ & $1 / \nu$ & $\gamma/\nu+2\beta/\nu$ & $2\beta'/\nu$ \\ 
    \hline
    2d Ising & 0.125 & 1.75 & 1 & 2 & 0.25 \\
    2d Ising (sim.) & 0.113(1) & 1.751(2) & 0.97(3) & 1.97 & 0.249(1) \\
    ABPs \cite{Siebert:2018} & 0.45 & 1.47 & 0.67 & 2.07 & - \\
    model I (hex.) & 0.1567(3) & 1.678(2) & 1.03(2) & 2.00 & 0.334(2) \\
    model I (sq.) & 0.1528(1) & 1.695(3) & 1.023(8)  & 2.00 & 0.327(4) \\
    model II (sq.) & 0.2208(1) & 1.649(1) & 1.021(7) & 2.10 & 0.391(6) \\
    \hline\hline
    \end{tabular}
    \caption{Comparison of critical exponents. The first three columns are the estimated values. The fourth column is the hyperscaling relation Eq.~\eqref{eq:hyper}, which is approximately obeyed by all models. The last column shows results for the alternative determination of $\beta'$ (cf. Sec.~\ref{sec:alt_bet}). Errors refer to statistical errors obtained by splitting respective data sets into three parts and calculating the standard error of the mean.}
    \label{tab:summary}
\end{table}

Based on our numerical results, we find the general conclusion from Ref.~\cite{Partridge:2019} that MIPS belongs to the 2d Ising universality class to be somewhat premature. Our results even cast some serious doubts on the weaker claim that model I exhibits 2d Ising behaviour. At this point we would like to emphasize that the numerical evidence presented in Ref.~\cite{Partridge:2019} is based on figures similar to our Figs.~\ref{fig:hlm_crex}(b) and (c) in which the slopes for the 2d Ising values were drawn on top of the simulation values suggesting excellent agreement. However, the authors neither provide values for $\gamma$ or $\nu$, nor did they mention the discrepancy for the exponent $\beta$.

Instead, we see mounting evidence that the critical behaviour for models exhibiting MIPS is at least to some degree model-dependent. Whether or not there is an underlying Ising universality or any universality at all, and to which extent deviations occur and why still remains an interesting and challenging question for simulations and theory alike.


\appendix
\section{Hyperscaling relation}
\label{sec:hyper}

The scaling relation Eq.~\eqref{eq:hyper} is typically derived from a free energy following arguments originally developed by Widom~\cite{Widom:1965a}. Since active matter is steadily driven away from thermal equilibrium, its behavior is not governed by such a free energy. However, the property of the free energy that is mostly exploited in deriving scaling laws is that of a \emph{generating function}, and some of the results can be transferred to non-equilibrium systems.

To this end, consider the cumulant generating function
\begin{equation}
    \phi(\tau,h) = \ln\sum_{\mathcal C} p(\mathcal C;\tau) e^{h\hat m(\mathcal C)}
\end{equation}
for the order parameter $\hat m=\hat m(\mathcal C)$ summing over all possible configurations $\mathcal C$ with probability $p(\mathcal C;\tau)$ depending on the control parameter $\tau$. The auxiliary field $h$ allows to obtain the average $m$ and susceptibility $\chi$ as
\begin{equation}
    m = \left.\pd{\phi}{h}\right|_{h=0} = \mean{\hat m}, \qquad \chi = \left.\pd{^2\phi}{h^2}\right|_{h=0} = \mean{\hat m^2} - m^2.
\end{equation}
A system with linear extend $l$ and correlation length $\xi$ in $d$ dimensions can be viewed as $n\simeq(l/\xi)^d$ independent systems, for which the joint probability $p(\mathcal C)=\prod_{i=1}^n p_\xi(\mathcal C_i)$ becomes a product of probabilities $p_\xi(\mathcal C';\tau)$ for the configuration in a smaller system with linear extend $\xi$. With $\hat m(\mathcal C)=\sum_{i=1}^n\hat m(\hat C_i)$ we have 
\begin{equation}
    \phi(\tau,h) \simeq \ln\left[\sum_{\mathcal C'}p_\xi(\mathcal C')e^{h\hat m(\mathcal C')}\right]^n \simeq (l/\xi)^d\tilde\phi(h/|\tau|^\Delta).
\end{equation}
In the second step we invoke the usual scaling hypothesis positing a scaling function $\tilde\phi(x)$ of the combined argument $h/|\tau|^\Delta$ with gap exponent $\Delta$. With $\xi\sim\tau^{-\nu}$ we thus find $m\sim\tau^{d\nu-\Delta}\sim\tau^\beta$ and $\chi\sim\tau^{d\nu-2\Delta}\sim\tau^{-\gamma}$. Eliminating $\Delta=d\nu-\beta$ leads to $\gamma+2\beta=d\nu$ [Eq.~\eqref{eq:hyper}]. Hence, this scaling relation only requires extensivity and homogeneity of the generating function, two properties that are not restricted to equilibrium.

\section{Critical slowing down}
\label{sec:slowing}

\begin{table}[t]
    \centering
    \begin{tabular}{c|c|c|c}
    \hline\hline
    model & run time & MC steps per individual run & number of individual runs \\ 
    \hline
    2d Ising & 15 days & ca. 1,230,000,000 & 20 \\
    model I (hex.) & 15 days & ca. 170,000,000 & 12 \\
    model I (sq.) & 15 days & ca. 175,000,000 & 12 \\
    model II (sq.) & 15 days & ca. 285,000,000 & 12 \\
    \hline\hline
    \end{tabular}
    \caption{Amount of simulation data for the biggest simulation box ($l=42$) at each respective critical point. A comparable amount of data is used for each of the other simulation points. The same run time is applied to the smaller simulation boxes, therefore more data is available for these systems. The difference in MC steps performed is a result of different computational effort for the particular models. The equilibration time is not included.}
    \label{tab:simdata}
\end{table}

\begin{figure}[ht]
  \centering
  \includegraphics[width=0.6\textwidth]{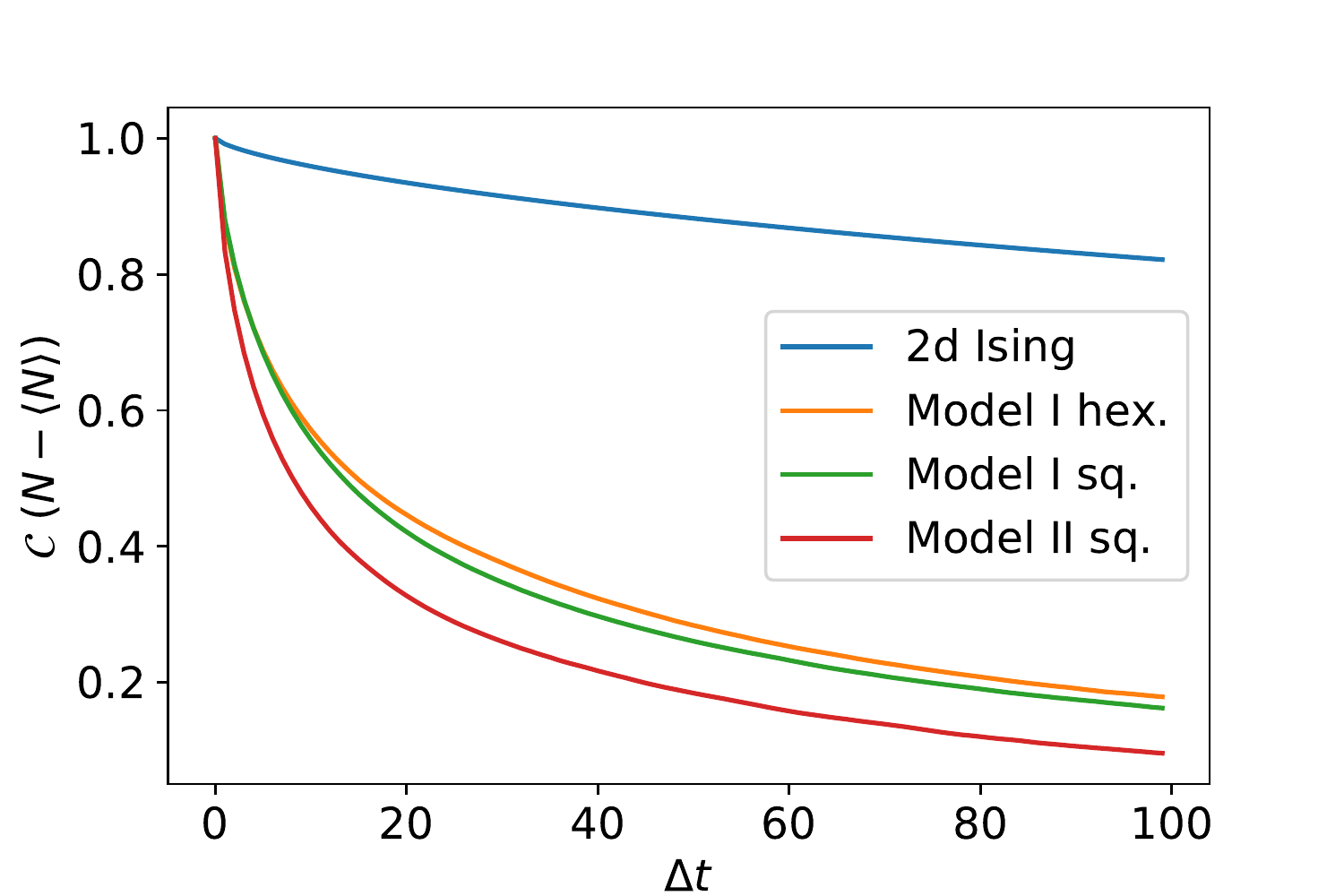}
  \caption{Autocorrelation of the density fluctuations $N-\langle N \rangle$ for the different models plotted for the largest system size of $l=42$ and at the respective critical point. Averaged over all individual subboxes and all simulation data available.}
  \label{fig:autocor}
\end{figure}

As mentioned above, the statistical quality of the data obtained for the active lattice models is way better than that obtained for the 2d Ising model, despite comparable simulation run time. We determine the autocorrelation function for $N- \langle N \rangle$ in order to estimate effects of critical slowing down. The strongest effects can be found for the biggest systems, therefore Fig.~\ref{fig:autocor} only shows the results for $l=42$ (simulation box of size $84\times 252$). Each model is evaluated at its respective critical point. The autocorrelation function is calculated for each individual subbox separately and then averaged over all subboxes and all simulation data obtained. The time lag $\Delta t$ is given in units of 1,000 Monte Carlo (MC) steps. For each MC step, each lattice site is on average picked once. So in total $\Delta t = 1$ is in this case equivalent to performing 21,168,000 simulation steps. Table.~\ref{tab:simdata} provides an overview of the total amount of simulation data evaluated for this work.

Fig.~\ref{fig:autocor} shows that the correlation for the 2d Ising model is quiet persistent, indicating critical slowing down. On the other hand, the correlations decay rather quickly for the active lattice models and are more or less gone after $\Delta t = 100$. Consequently, critical slowing down is not an issue for the active lattice models. The main reason for these differences are the acceptance schemes in the models. For the Ising model, the metropolis criterion has to be fulfilled for a spin-swap. In comparison, each move to a free lattice site is accepted for the active lattice models and there is no interaction except for hard repulsion.


\begin{acknowledgements}
We thank C. Maggi and K. Binder for illuminating discussions. We gratefully acknowledge financial support by the Deutsche Forschungsgemeinschaft within priority program SPP 1726 (Grants No. SP1382/3-2 and No. VI 237/5-2). ZDV Mainz is acknowledged for computing time on the MOGON supercomputers.
\end{acknowledgements}

{\noindent\small\textbf{Author contribution statement} PV and TS designed the research. FD wrote and performed the simulations and analyzed the data. All authors contributed to writing the manuscript.}


\end{document}